\documentclass[a4paper,10pt]{article}

\usepackage{amstext,amsmath,amssymb}
\usepackage[dvips]{graphicx}
\usepackage{latexsym}

\newcommand{\R}{\mathbb{R}}
\DeclareMathOperator{\tr}{tr}

\def\be{\begin{equation}}
\def\ee{\end{equation}}
\def\bea{\begin{eqnarray}}
\def\eea{\end{eqnarray}}
\def\nn{\nonumber}

\newcommand{\lalg}[1]{\mathfrak{#1}}

\newcommand{\SO}{\mathrm{SO}}

\newcommand{\Ad}{\mathrm{Ad}}

\newcommand{\so}{\lalg{so}}

\newcommand{\p}{\mathcal{P}}
\newcommand{\s}{\mathcal{S}}
\newcommand{\A}{\mathbf{A}}
\newcommand{\F}{\mathbf{F}_{\mathbf{A}}}
\newcommand{\U}{\mathbf{u}}
\newcommand{\W}{\mathbf{w}}

\begin{document}

\title{On gravitational defects, particles and strings}

\author{{\bf Winston J. Fairbairn} \footnote{{winston.fairbairn@nottingham.ac.uk}} \\
School of Mathematical Sciences \\ University of Nottingham \\ University Park \\
Nottingham \\ NG7 2RD, UK}

\maketitle

\vspace{6mm}

\begin{abstract}

\noindent We study the inclusion of point and string matter in the deSitter gauge theory, or MacDowell-Mansouri formulation of four dimensional gravity. We proceed by locally breaking the gauge symmetries of general relativity along worldlines and worldsheets embedded in the spacetime manifold. Restoring full gauge invariance introduces 
new dynamical fields which describe the dynamics of spinning matter coupled to gravity.

We discuss the physical interpretation of the obtained formalism by studying the flat limit and the spinless case on arbitrary backgrounds. It turns out that the worldline action describes a massive spinning particle, while the worldsheet action contains the Nambu-Goto string augmented with spinning contributions.

Finally, we study the gravity/matter variational problem and conclude by discussing potential applications of the formalism to the inclusion of the Nambu-Goto string in spinfoam models of four dimensional quantum gravity.

\end{abstract}

\newpage

\tableofcontents


\section{Introduction}

Our common quantum relativistic understanding of matter in terms of finite dimensional, irreducible representations of the Poincar\'e algebra is a very rough approximation of reality. This description is tied to the isometries of the flat, Minkowski solution to general relativity and yields a good approximation only in very weak gravitational fields, like for instance, in our particle accelerators where the successes of quantum field theory have been crowned.

In a fundamental theory of Nature, one cannot expect this approximation to be valid since in the early, Planckian universe, spacetime is undoubtedly not flat. Accordingly, the search of the fundamental structure of matter is tied to non-trivial, and certainly quantum configurations of the gravitational field. In turn, a complete theory of quantum gravity will have to incorporate a precise description of the degrees of freedom of matter.

As a first step, it seems therefore natural to look for an understanding of matter which does not rely on a particular fixed background geometry at the classical level. This will automatically render the formulation compatible with  non-perturbative attempts to the quantisation of gravity which cannot, consistently, rely on a fixed, background metric structure.

A very old and appealing idea consists in considering the Einstein equations as defining the notion of matter. In other words, to consider matter as particular, possibly singular, configurations of the gravitational field. In this framework, we are reversing the standard picture where matter is defined on flat spacetime and then tentatively extended to other solutions of general relativity. Here, we are starting from the gravitational perspective, without selecting a preferred solution, and deriving matter from the geometry of spacetime.
Obviously, this formulation should reproduce the standard properties of matter in the flat limit, but will also select
a preferred formulation from the gravitational perspective. For example, such a reversed approach has recently led to conceptually and technically strong results regarding the coupling of matter to three dimensional quantum gravity \cite{effqg}, \cite{karim}. 

The concrete implementation of this procedure relies on a the gauge symmetries of gravity, that is, diffeomorphisms and local Lorentz transformations. The idea is to locally break these symmetries by introducing fixed membranes of various dimensions embedded in the spacetime manifold. Restoring the full gauge invariance transforms the would-be gauge parameters into dynamical variables living on the membranes. These new fields describe the matter degrees of freedom; the diffeomorphism and Lorentz gauge symmetries being respectively related to the embedding, and the momentum and spin of the membrane.

The detailed account of the above derivation in the deSitter gauge theory formulation of gravity is the content of the first section of this paper. This covers the particle case derived in \cite{artem} and the string case which is new. In section two, we study the physical interpretation of the obtained formalism by making contact with the description of matter in flat space in terms of pseudo-classical variables \cite{Bal1}, \cite{Bal2}, \cite{Stern}. 
In this context, we compute the invariants associated to a particle and to a string to unravel the physical picture emerging from the formalism.
Finally, section three is devoted to the variational problem of the actions of section one. We review the results of \cite{artem} for the particle, and find similar equations of motion for the string. We derive the corresponding Einstein equations and discuss some solutions for the spinless string case.
Finally, we discuss some perspectives concerning the inclusion of the Nambu-Goto string in spinfoam models of four-dimensional quantum gravity.

\section{Gauge defects in general relativity}

We firstly recall the basics of the McDowell-Mansouri formulation of four-dimensional general relativity (GR) with positive cosmological constant. Then, we include matter by locally breaking the symmetries of the theory.

\subsection{deSitter gauge theory formulation of gravity}

The idea that first order GR with zero cosmological constant is related to a gauge theory of the Poincar\'e group has been investigated for the last fifty years following the pioneering works of Utiyama \cite{Utiyama}, Kibble \cite{Kibble} and Sciama \cite{Sciama}. In this framework, the fundamental fields of first order GR, the coframe $e$ and metric connection $A$, can be combined into a single Cartan connection taking value in the Poincar\'e Lie algebra. In fact, it turns out that this more unified description, relying on the notion of Cartan geometry, can be extended to all values of the cosmological constant. The idea is to construct first order GR from a gauge theory with symmetry group the isometry group associated to the maximally symmetric solution of the corresponding Einstein equations, that is, the Poincar\'e, deSitter and anti-deSitter groups in the $\Lambda=0$, $\Lambda>0$ and $\Lambda<0$ cases respectively. 
For a detailed account on the relation between gravity and Cartan geometry, see \cite{wise}.

However, unlike three-dimensional gravity, GR in four dimensions is only invariant under the Lorentz subgroup of the above larger isometry groups. For instance, in the $\Lambda=0$ case, the Einstein-Cartan Lagrangian is not invariant under local translations and thus is not Poincar\'e invariant. Hence, four-dimensional gravity is not a true gauge theory but rather a broken phase of a gauge theory based on a larger isometry group than the Lorentz group.
This symmetry breaking can be described in terms of Higgs fields in the Cartan gauge theories where GR is obtained by choosing the appropriate Lorentz-invariant ground state. See \cite{Leclerc} and references therein for details.

This gauge theory formalism of gravity has a substantial geometrical interpretation, and has shown to be powerfull in the context of three-dimensional quantum gravity. For our purposes, namely the introduction of matter, it will be very convenient, allthough not necessary, to pack the co-frame $e$ and the connection $A$ in a single object and to work within this formalism. 
In this paper, we will essentially work in the $\Lambda>0$ case, supported by experimental evidence, in both Lorentzian and Riemannian signatures, keeping track of the $\Lambda=0$ situation as a limiting case. 
Hence, we will be interested in the isometry group $G=\SO(\eta_5)$ of a flat {\itshape five dimensional} metric $\eta_5=(\sigma^2,+,+,+,+)$, with $\sigma=i$ (resp. $\sigma=1$) in Lorentzian (resp. Riemannian) signatures, that is, the deSitter group $G=\SO(4,1)$ (resp. the rotation group $G=\SO(5)$).

\subsubsection{The deSitter algebra}

The deSitter algebra $\so(\eta_5)$ is a real ten-dimensional semi-simple Lie algebra.  Let $(\sigma_{AB})_{A < B=0,...,4}$, $\sigma_{AB} = - \sigma_{BA}$, denote a basis of generators, and $(\pi, V_{\eta_5})$ be the vector representation of $\so(\eta_5)$. In this representation, the generators are given by the ten, five by five matrices with matrix elements $\pi(\sigma_{AB})^I_{\; J} = \delta^I_{5 \, A} \eta_{5 \, JB} - \eta_{5 \, AJ} \delta^I_{5 \, B}$, $I,J=0,...,4$. Consider the basis relabelling leading to the $6+4$ decomposition of $\so(\eta_5)$:
\be
\label{split}
\so(\eta_5) = W \oplus V,
\ee
where 
$$
W = \R \{ \sigma_{ab}\}_{a<b=0,...,3}, \;\;\;\; \mbox{and} \;\;\;\; V = \R \{ P_{a}\}_{a=0,...,3}, 
$$
with $P_a = l^{-1} \, \sigma_{a4}$. The rescalling\footnote{Note that the components of the elements of $\so(\eta_5)$ in the above basis are rescalled accordingly, i.e., all $X$ in $\so(\eta_5)$ is written $X= (1/2) X^{AB} \sigma_{AB} = (1/2) X^{ab} \sigma_{ab} + X^a P_a$, with $X^a = l X^{a4}$ acquiring the dimension of a length.} 
of the generators involves a factor $l$ with dimension of a length. It is related to the cosmological constant $\Lambda>0$ by requiring that deSitter space with radius $l$ is a solution to the vacuum Einstein equations with positive cosmological constant $\Lambda$. This leads to the relation $l^2 = 3/\Lambda$. 

The matrix elements of the five by five matrices $\pi(\sigma_{ab})$ and $\pi(P_a)$ are read out of the vector representation of the deSitter generators $\sigma_{AB}$.

In this presentation, the Lie algebra structure of $\so(\eta_5)$ is given by
\bea
\label{deSitter}
&& [\sigma_{ab},\sigma_{cd}]=\eta_{ad} \sigma_{bc}+\eta_{bc} \sigma_{ad}-\eta_{bd} \sigma_{ac}-\eta_{ac} \sigma_{bd} \nn \\
&& [\sigma_{ab},P_c] = \eta_{bc} P_a - \eta_{ac} P_b \nn \\
&& [P_a,P_b] = - \frac{1}{l^2} \sigma_{ab}.
\eea
In fact, we have constructed a family of semi-simple Lie algebras parametrised by the real number $l$. For finite $l$, all these algebras are isomorphic as Lie algebras and the structure constants approach a well defined limit for $l \rightarrow \infty$, when the deSitter radius becomes large, that is, when the cosmological constant goes to zero. However, the Lie algebra recovered in this limit is no longer isomorphic to the Lie algebras obtained for finite $l$. The algebra obtained by this contraction is the Poincar\'e algebra (or the Euclidean algebra) $i\so(\eta)$. 

The above Lie algebra structure shows that the split \eqref{split} only occurs at the level of vector spaces. However, the subvector space $W$ is in fact a subalgebra generating an isometry subalgebra $\so(\eta)$ in $\so(\eta_5)$, where $\eta = (\sigma^2,+,+,+)$ is the diagonal form of a Lorentzian (resp. Riemannian) metric $g$ on $M$, i.e., $\so(\eta)=\so(3,1)$ (resp. $\so(\eta) = \so(4)$).
Note that allthough the split \eqref{split} is not deSitter invariant, it is preserved under the adjoint action of the Lorentz subgroup; the underlying Cartan geometry is reductive.
Concerning the subspace $V$, the commutation relations \eqref{deSitter} show that it is not a subalgebra but is isomorphic as a vector space to the vector representation $V_{\eta}=\R^{3,1}$ (resp. $V_{\eta}=\R^4$) of $\so(\eta)$, by dimensional considerations.

Therefore, there exists an inclusion map
\be
\label{emb}
\iota : \so(\eta) \oplus V_{\eta} \rightarrow W \oplus V,  
\ee
embedding $\so(\eta)$ (resp. $V_{\eta}$) into the deSitter algebra $\so(\eta_5)$ as a Lie subalgebra (resp. as subvector space).

Using this isomorphism, we can relate the adjoint action of the isometry subgroup associated to the split \eqref{split} on the translational part $V$ of the deSitter algebra $\so(\eta_5)$ and the action of an isometry in $V_{\eta}$. A Lorentz transformation (or a rotation) $\Lambda$ in $V_{\eta}$ is related to the adjoint action on $\iota(V_{\eta})$ by the relation $\iota(\Lambda v) = \Lambda \iota(v) \Lambda^{-1}$, forall $v$ in $V_{\eta}$, where on the right hand side $\Lambda$ is an element of the subgroup $\SO(\eta) \subset \SO(\eta_5)$. In terms of the chosen basis of $V$, the relation yields $\Lambda P_a \Lambda^{-1} = \Lambda^b_{\; a} P_b$.

We now endow the deSitter algebra with a metric structure. 
Let $\langle , \rangle = -\frac{1}{2} \tr$ be a non-degenerate, symmetric bilinear form on $\so(\eta_5)$, constructed from the trace `$\tr$' in the vector representation. This metric is invariant under the adjoint action of the deSitter group 
\be
\label{invariance}
\langle \W_1 , \W_2 \rangle = \langle \Ad_{g} (\W_1) , \Ad_{g} (\W_2) \rangle,
\ee
with $\Ad_{g} (\W):= g \W g^{-1}$, forall $\W_i$ in $\so(\eta_5)$ and $g$ in $\SO(\eta_5)$, and is accordingly an invariant on each orbit of the adjoint action of the Lorentz subgroup $\SO(\eta)$ on the deSitter Lie algebra $\so(\eta_5)$. Since the deSitter algebra is semi-simple, the above metric is necessarily proportional to the Killing form.

This bilinear form is defined by the evaluations
\bea
\label{bilin}
\langle \sigma_{ab}, \, \sigma_{cd} \rangle &=& \eta_{ac} \eta_{bd} - \eta_{ad} \eta_{bc}  \nn \\
\langle P_a, \, P_b \rangle &=& \frac{1}{l^2} \eta_{ab} \nn \\
\langle \sigma_{ab}, \, P_c \rangle &=& 0.
\eea 
This shows that the split \eqref{split} is in fact an orthogonal decomposition of the deSitter algebra in the above bilinear form, i.e., $V=W^{\bot}$ and that the isomorphism $W \simeq \so(\eta)$ is an isometry with the isometry algebra equipped with its standart bilinear form, while the metrics on $V_{\eta}$ and $V$ are equivalent up to rescaling.

Note that in the $l \rightarrow \infty$ contraction, the above metric becomes degenerate when restricted to the translational part. This is due to the non-semi-simplicity of the Poincar\'e (or Euclidean) algebra and this bilinear form can only be used in the finite radius $l$ case. In the non-semi-simple limit, one has to work with a different bilinear form  obtained by relaxing the full Poincar\'e $\Ad$-invariance requirement. In this case, exploiting the semi-direct product structure, one can obtain a perfectly well defined, non-degenerate, $\SO(\eta)$-invariant, symmetric bilinear form $( , )$ on $i\so(\eta)$ defined by the evaluation
\be
\label{Poincare}
( \U_1 , \U_2 ) = (w_1 \oplus v_1 , w_2 \oplus v_2) = - \frac{1}{2} \tr \, w_1 w_2 + \eta(v_1 , v_2),
\ee
forall $\U_i = w_i \oplus v_i$ in the Poincar\'e algebra $i \so(\eta) = \so(\eta) \oplus V_{\eta}$.

\subsubsection{deSitter gauge theory formalism: action and symmetries} 

\paragraph{Action.}
The dynamical fields of the first order formulation of gravity on a four-dimensional, oriented, differential manifold $M$ are defined by the pair ($e,A$) where $e$ is the soldering form on $M$ and $A$ is (the pull-back to $M$ by local sections of) a metric connection on the bundle of $\eta$-orthonormal frames, that is, the principal bundle over $M$ with structure group $\SO(\eta)$.
The direct sum decomposition \eqref{split} (together with the isomorphism \eqref{emb}) can be exploited to combine the co-frame $e$ and the metric connection $A$ in a $\so(\eta_5)$ Cartan connection 
\be
{\mathbf{A}} = A \oplus e. 
\ee
Computing the curvature of the Cartan connection yields a Lorentz and a pseudo-translational component
\be
F_{\mathbf{A}} = (F_A - \frac{\Lambda}{3} e \wedge e) \oplus d_A e,
\ee
where $F_A = dA + \frac{1}{2}[A,A]$ is the curvature of the connection $A$, and $d_A$ is the covariant derivative with respect to the connection $A$.

Constructing a suitable gravitational Lagrangian out the above field strength requires an expression with no metric dependence. If one further requires to have a fully deSitter invariant quantity, the Lagrangian must also be quadratic in the field strength. Combining these two requirements leads to a unique possible Lagrangian which turns out to have trivial variations, i.e., corresponds to a topological theory. But gravity is not a gauge theory of the deSitter group, it is only invariant under local Lorentz transformations. It is precisely the breaking of the deSitter invariance down to Lorentz invariance that introduces the local degrees of freedom of gravity into the formalism. To achieve this symmetry breaking, we introduce the projection map $\phi: W \oplus V \rightarrow W$ associated to the orthogonal decomposition \eqref{split}. The McDowell-Mansouri (MM) action \cite{mm} then follows
\be
\label{MM}
S_{\mbox{{\tiny MM}}}[\mathbf{A}] = \alpha \int_M \langle * \phi[F_{\mathbf{A}}] \wedge \phi[F_{\mathbf{A}}] \rangle.
\ee
Here, the star $* : \Omega^2(V) \rightarrow \Omega^2(V)$ denotes the internal Hodge operator acting on the exterior algebra over $V$, i.e., $*(X)_{ab} = \frac{1}{2} \epsilon_{ab}^{\;\;\; cd} X_{cd}$, forall $X$ in $ \Omega^2(V)$. We are implicitly using the isomorphism $\Omega^2(V) \simeq \so(\eta)$ between the space of two-forms over $V \simeq V_{\eta}$ and the isometry algebra. 
The coupling constant $\alpha$ is related to the Newton constant $G$ by $\alpha = 3 / (16 \pi G \Lambda)$.

It is immediate to see that the above action 
reduces to the Einstein-Cartan action with cosmological constant
\be
\label{pureaction}
S_{\mbox{{\tiny GR}}}[e,A] = - \frac{1}{16 \pi G} \int_M \tr \, * e \wedge e \wedge ( F_A \, - \frac{\Lambda}{6} e \wedge e ),
\ee
augmented with a topological term:
\be
S_{\mbox{{\tiny MM}}}[e,A] = - S_{\mbox{{\tiny GR}}} + \alpha \int_M \tr * F_A \wedge F_A .
\ee
The second term is the integral of the Euler class which has trivial variation because of the Bianchi identity $d_A F_A=0$. Thus, the de-Sitter gauge theory defined by the action \eqref{MM} is classically equivalent to GR.

Note that the zero cosmological constant limit of the Mac-Dowell-Mansouri action \eqref{MM} is not GR with zero cosmological constant, but rather a topological field theory. 
To obtain GR from the Cartan perspective, one needs to contract the deSitter algebra to the Poincar\'e algebra, change the bilinear form and start with a different Lagrangian; the relation of the gauge theory Lagrangians associated to different values of the cosmological constant is an open question.

\paragraph{Symmetries.}
The theory is invariant under two sets of symmetries. First, 
it is invariant under the action of the (connected component of the) diffeomorphism group Diff($M$). 
If $\phi:M \rightarrow M$ is an diffeomorphism of $M$, the action is invariant under the corresponding action on the gauge field 
\be
\label{diff}
\A \mapsto \phi^{-1*} \A,
\ee
where $\phi^*$ is the associated pull-back map acting on forms. Note that this transformation acts on the space of fields, even if it is induced by an action on the manifold. 
Next, as remarked above, the theory is invariant under local Lorentz transformations:
\be
\mathbf{A} \mapsto - d g g^{-1} + g \mathbf{A} g^{-1},
\ee
where 
$g$ is a smooth map from $M$ to the Lorentz subgroup $\SO(\eta) \subset \SO(\eta_5)$ associated to the split \eqref{split}.

\newpage

To introduce matter, we now follow the ideas recently developed in the BF theory context in three \cite{Freidel1} and four dimensions \cite{BP}, \cite{AW}, \cite{Ale}, \cite{artem}. The idea is to locally break the diffeomorphism and gauge symmetry of GR by fixing sub-manifolds of $M$ and integrating against them suitable powers of the fields $e$ and $A$, i.e., of the Cartan connection $\mathbf{A}$, projected along preferred directions in the appropriate tensor power of $\so(\eta_{5})$. Then, one can promote the would-be gauge parameters to local degrees of freedom. 

\subsection{Inclusion of defects}

\subsubsection{Worldlines}

The simplest way of proceeding consists in introducing a fixed, non-dynamical, one-dimensional submanifold $\gamma \subset M$, defined by the fixed embedding $x: \R \rightarrow M$; $\tau \mapsto x(\tau)$, and a fixed constant element $\mathbf{u}^*$ in the dual of the deSitter algebra $\so(\eta_5)^*$. We note $\mathbf{u}=w \oplus v$ the element of the deSitter algebra associated to $\mathbf{u}^*$ via the Killing form, i.e., $\U^*(\mathbf{w}) = \langle \U , \mathbf{w} \rangle$ for any $\mathbf{w}$ in $\so(\eta_5)$. Consider adding to the action \eqref{MM} the following symmetry breaking functional of the gauge field
\be
S[\mathbf{A}] = \int_{\gamma} \langle \mathbf{u}, \mathbf{A}  \rangle.
\ee
The above term breaks the symmetry of the theory at the location of the curve $\gamma$; the diffeomorphism gauge symmetry is broken down to a symmetry under the diffeomorphisms keeping $\gamma$ fixed, and the Lorentz gauge symmetry is broken down to an invariance under the Lorentz subgroup leaving the fixed internal vector $\mathbf{u}$ invariant. Note that the above term is nevertheless invariant under reparametrisation of the curve $\gamma$.

We now wish to restore the gauge invariance of the theory. We proceed by performing an {\em inverse} gauge transformation on the dynamical gauge field $\mathbf{A}$, that is, an operation on the space of fields such that its composition with gauge transformation compensates to the identity. This operation will obviously only affect the symmetry breaking term. Firstly, we perform an inverse diffeomorphism gauge transformation $\mathbf{A} \mapsto \phi^* \mathbf{A}$, with gauge parameter $\phi$. The gauge transformed symmetry breaking term $S[\phi^* \mathbf{A}]$ now also depends on (the restriction $X:\gamma \rightarrow M$ to $\gamma$ of) the gauge parameter $\phi$ and becomes
\bea
\label{partial}
S[\mathbf{A};X] &=& \int_{\gamma} \langle \mathbf{u}, X^* \mathbf{A}  \rangle \nn \\
                &=& \int_{X(\gamma)} \langle \mathbf{u}, \mathbf{A}  \rangle,
\eea
where, in a local coordinate patch, $X^* \A = \A_{\mu}(X(\tau)) \partial_{\tau} X^{\mu}(\tau) d \tau$.
By doing so, we have introduced a new variable in the theory; the pull back to $\gamma$ of the would-be gauge parameter. Now, $X$ in Diff($\gamma,M$) is promoted to a dynamical variable\footnote{More precisely, it is the composition $X \circ x : \R \rightarrow M$ of the diffeomorphism $X$ with the fixed embedding $x$ that we are interpreting as the dynamical worldline of the particle.}. As such, we have to specify its transformation properties. It is required to transform trivially under local Lorentz transformation and as
\be
\label{diffgauge}
\forall \phi \in \mbox{Diff}(M), \;\;\;\; X \mapsto \phi \circ X,
\ee
under spacetime diffeomorphism. Then, the very diffeomorphism symmetry lost by the introduction of the fixed curve $\gamma$, that is, the invariance under diffeomorphisms shifting the curve, is restored in the (partial) symmetry breaking term \eqref{partial} by rendering the gauge parameter $X$, or equivalently, the worldline $X(\gamma)$, dynamical.

We can now proceed in an analogous way to restore the Lorentz gauge invariance of the theory. We perform an inverse finite gauge transformation  $\mathbf{A} \mapsto g^{-1} d_{\mathbf{A}} g$ labelled by the parameter $g$ in the (partial) symmetry breaking term \eqref{partial}. 
This procedure introduces a new field $\Lambda : \gamma \rightarrow \SO(\eta)$ in the theory, the restriction to $\gamma$ of the gauge parameter $g$. This field is promoted to a dynamical variable transforming trivially under spacetime diffeomorphisms and as
\be
\label{gauge}
\forall g \in C^{\infty}(M,\SO(\eta)), \;\;\;\; \Lambda \mapsto g \circ \Lambda,
\ee
under Lorentz gauge transformations. Then, introducing the deSitter-valued one-form
\be
\label{one-form}
\Omega_{\mathbf{A}} = \Lambda^{-1} d \Lambda + \Lambda^{-1} \mathbf{A} \Lambda,                  
\ee
we are led to the dynamical, gauge invariant action \cite{artem}:
\be
\label{particleMM}
S_{\mbox{{\tiny particle}}}[\A;X,\Lambda] = \int_{X(\gamma)} \langle \mathbf{u}, \Omega_{\mathbf{A}} \rangle. 
\ee
Exploiting the orthogonal decomposition \eqref{split}, it is immediate to recast the above action in the form obtained in \cite{Bal1} by Balachandran and collaborators :
\be
\label{particleaction}
S_{\mbox{{\tiny particle}}}[e,A;X,\Lambda] =  \int_{X(\gamma)} \langle v , \Lambda^{-1} e \rangle + \langle w , \Lambda^{-1} d_A \Lambda \rangle  
\ee
where we have used the decomposition in reductive components of the deSitter one form $\Omega_{\mathbf{A}}$:
$$
\Omega_{\mathbf{A}}= \Lambda^{-1} d_A \Lambda \oplus \Lambda^{-1}e,
$$
with, in the translational part, the matrix $\Lambda$ acting on the algebra indices $a,b$. 
As we will see, for specific values of the elements $w$ and $v$, the first term becomes the standart first order particle action while the second term is a Wess-Zumino term describing a pseudo-classical spin. 

\subsubsection{Worldsheets}

Similarly, we can break the gauge invariance of GR along two-surfaces. 
A natural object to integrate along a two-dimensional submanifold of $M$ is the second exterior power of the Cartan connection $\mathbf{A}$. This object takes value in the antisymmetric part of $\so(\eta_5) \otimes \so(\eta_5)$. 

The above tensor square admits a decomposition induced by the split \eqref{split} given by
\be
\label{split2}
\so(\eta_5) \otimes \so(\eta_5) = (W \otimes W) \oplus (W \otimes V) \oplus (V \otimes W) \oplus (V \otimes V).
\ee
This vector space can be endowed with a Lie algebra structure \cite{Ellis}, but for our purposes it will be sufficient to consider this space at the level of metric vector spaces. Note however that the antisymmetric part of $\so(\eta_5) \otimes \so(\eta_5)$ contains the vector space $\Omega^2(V) = \R \{ P_a \wedge P_b \}_{a < b}$ isomorphic to $\Omega^2(V_{\eta})$ and to the Lorentz algebra $\so(\eta)$. 

To endow the tensor square with a metric structure, we consider
the bilinear form $\langle , \rangle := \frac{1}{2} \langle , \rangle^{\otimes 2}$ constructed from the tensor product of the bilinear form $\langle , \rangle$ with itself. This non-degenerate, symmetric bilinear form on $\so(\eta_5)^{\otimes 2}$ is invariant under the diagonal adjoint action of the deSitter group:
\be
\label{invariance1}
\langle [\Ad_{g} \otimes \Ad_{g} ](\mathbf{w}_1), [\Ad_{g} \otimes \Ad_{g} ](\mathbf{w}_2) \rangle = \langle \mathbf{w}_1, \mathbf{w}_2 \rangle,
\ee
forall $\mathbf{w}_i$ in $\so(\eta_5)^{\otimes 2}$, and  $g$ in $\SO(\eta_5)$. Accordingly, it is also an invariant on each orbit of the $\SO(\eta)$ subgroup action.

The explicit evaluations of this bilinear form on the basis elements associated to the decomposition \eqref{split2} read
\bea
\label{bilin1}
\langle \sigma_{ab} \otimes \sigma_{ef}, \, \sigma_{cd} \otimes \sigma_{gh} \rangle &=& \frac{1}{2} (\eta_{ac} \eta_{bd} - \eta_{ad} \eta_{bc}) (\eta_{eg} \eta_{fh} - \eta_{eh} \eta_{fg}) \nn \\
\langle \sigma_{ab} \otimes  P_e, \, \sigma_{cd} \otimes  P_f \rangle &=& \frac{1}{2l^2} (\eta_{ac} \eta_{bd} - \eta_{ad} \eta_{bc}) \eta_{ef} \nn \\
\langle P_a \otimes \sigma_{cd}, \, P_b \otimes \sigma_{ef} \rangle &=& \frac{1}{2l^2} \eta_{ab}(\eta_{ce} \eta_{df} - \eta_{cf} \eta_{de}) \nn \\ 
\langle P_a \otimes P_c, \, P_b \otimes P_d \rangle &=& \frac{1}{2l^4} \eta_{ab} \eta_{cd},
\eea 
with vanishing of all other evaluations.
Note that when restricted to $\Omega^2(V)$, this metric reduces to the standart bilinear form on $\so(\eta)$, up to rescalling:
$$
\langle P_a \wedge P_c, \, P_b \wedge P_d \rangle = \frac{1}{l^4} (\eta_{ab} \eta_{cd} - \eta_{ad} \eta_{bc}).
$$
Armed with this metric structure, we can now define the appropriate symmetry breaking term. 

We firstly introduce a fixed, non-dynamical, compact and closed two-surface $\Sigma \subset M$ defined by the fixed embedding $x: \R^2 \rightarrow M$; $\tau \mapsto x(\tau)$, with $\tau=(\tau_{\alpha})$, $\alpha=0,1$. We then fix a constant element $\mathbf{w}^*$ in the antisymmetric part of $\so(\eta_5)^* \otimes \so(\eta_5)^*$, associated to $\W$ in $\so(\eta_5) \otimes \so(\eta_5)$ via the bilinear form $\langle , \rangle$. 
In what follows, we will use the symbolic notation $\W=\mathbf{u}_1 \otimes \mathbf{u}_2$, with $\mathbf{u}_i=w_i \oplus v_i$ and antisymmetrised tensor product, to keep track of the decomposition of the element $\W$ according to the basis associated to \eqref{split2}.

Consider adding to the gravitational action \eqref{MM} the following symmetry breaking functional of the gauge field:
\be
S[\mathbf{A}] = \int_{\Sigma} \langle \mathbf{w}, \mathbf{A} \wedge \mathbf{A} \rangle.
\ee
This term breaks the diffeomorphism symmetry down to the diffeomorphisms keeping the surface $\Sigma$ fixed, and the Lorentz gauge invariance is broken down to an invariance under the Lorentz subgroup keeping the internal tensor $\mathbf{w}$ fixed under the diagonal adjoint action. Note that the symmetry breaking term is nevertheless invariant under reparametrisation of the worldsheet $\Sigma$.
As for the particle, we can now restore the gauge invariance of the theory by performing an inverse gauge transformation on the gauge field $\mathbf{A}$. 

We start by an inverse diffeomorphism gauge transformation labelled by $\phi$. The bulk action is invariant while the symmetry breaking term is shifted. Noting $X: \Sigma \rightarrow M$ the restriction to $\Sigma$ of $\phi$, it becomes
\bea
S[\mathbf{A};X] &=& \int_{\Sigma} \langle \mathbf{w}, X^*(\mathbf{A} \wedge \mathbf{A}) \rangle \nn \\
                &=& \int_{X(\Sigma)} \langle \mathbf{w}, \mathbf{A} \wedge \mathbf{A} \rangle
\eea
where, in a local coordinate patch,  
$$
X^*[\mathbf{A} \wedge \mathbf{A}] = \A_{\mu}(X(\tau)) \A_{\nu}(X(\tau)) \, \epsilon^{\alpha \beta} \, \partial_{\alpha} X^{\mu}(\tau) \, \partial_{\beta} X^{\nu}(\tau) \, d^2 \tau,
$$ 
with $\epsilon$ the two dimensional totally antisymmetric tensor normalised by $\epsilon^{01}=1$. Once that $X$ is promoted to a dynamical variable, transforming as specified in \eqref{diffgauge}, the above term becomes invariant under the action of Diff($M$).

Next, we perform an inverse finite gauge transformation with gauge parameter $g$ on the gauge field $\mathbf{A}$. 
The procedure introduces a new field $\Lambda:\Sigma \rightarrow \SO(\eta)$ in the theory, the restriction to $\Sigma$ of the gauge parameter $g$. This variable is then promoted to a dynamical field, transforming as in \eqref{gauge}, and the symmetry breaking term becomes a gauge invariant, dynamical action 
\bea
\label{stringMM}
S_{\mbox{{\tiny string}}}[\A;X,\Lambda] &=& \int_{X(\Sigma)} \langle \mathbf{w}, \Omega_{\A} \wedge \Omega_{\A} \rangle,
\eea
with the deSitter valued one-form $\Omega_{\A}$ defined as for the particle.

The introduction of the diffeomorphism $X$ in Diff($\Sigma,M$) has rendered the embedding of the string $\Sigma$ dynamical, thus restoring full diffeomorphism invariance, and the introduction of the group valued map $\Lambda : \Sigma \rightarrow \SO(\eta)$ has made the theory fully Lorentz gauge invariant. Hence,
the final action \eqref{stringMM} is invariant under spacetime diffeomorphisms, local Lorentz transformations and also under worldsheet diffeomorphisms.

It is important to stress that both the particle \eqref{particleMM} and string \eqref{stringMM} actions can be defined in the zero cosmological constant limit, by contracting the deSitter algebra to the Poincar\'e algebra, and respectively replacing the bilinear forms $\langle, \rangle$ and $\langle, \rangle^{\otimes}$, by $(,)$ and $( , )^{\otimes 2}$, with the Poincar\'e bilinear form $(,)$ defined in \eqref{Poincare}.

Using the decomposition \eqref{split2} to make the dependence on the co-frame $e$ and the Lorentz connection $A$ more transparent, 
one can see that the above action contains all gauge invariant, two-form combinations\footnote{Up to multiplication by the internal Hodge operator.} of $e$ and $A$
\bea
\label{stringaction}
S_{\mbox{{\tiny string}}}[e,A;X,\Lambda] &=& \int_{X(\Sigma)} \langle v_1 \otimes v_2, \, \Lambda^{-1} e \wedge \Lambda^{-1} e \rangle \nn \\
&& + 2 \int_{X(\Sigma)} \langle v_1 \otimes w_2, \, \Lambda^{-1} e \wedge \Lambda^{-1} d_A \Lambda \rangle \nn \\
&& + \int_{X(\Sigma)} \langle w_1 \otimes w_2, \, \Lambda^{-1} d_{A} \Lambda \wedge \Lambda^{-1} d_{A} \Lambda \rangle.
\eea
As we are about to see, the above actions describes a general spinning string on a non-trivial background spacetime. 

In the form given above \eqref{stringaction}, the first term appears to be the generalisation to curved spacetimes of the type of formulation of string theory considered by Balachandran and collaborators in \cite{Bal2}. For specific choices of the element $v_1 \otimes v_2$, this term describes a Nambu-Goto, null or tachionic string. The generalisation obtained by adding the second and third terms were considered in the flat case by Stern et al. \cite{Stern}, \cite{Stern1} and describe spinning contributions for the string.

We now study the precise point of contact between the symmetry breaking procedure followed here and this particular formulation of matter. This will help clarify the physical interpretation of the framework, since it is not immediate to grasp the physical picture underlying this description of matter. Indeed, the theories of gravitating particles or strings
considered above are described by the diffeomorphism and Lorentz gauge parameters of gravity pulled-back to the worldline or worldsheet, that is, a pair of smooth fields $(X,\Lambda)$ having support on the worldline or worldsheet and taking value in the spacetime manifold $M$ and the isometry group $\SO(\eta)$ respectively. How do these variables relate to standart notions such as position, momentum, spin?
While the variable $X$ can easily be interpreted as describing the embedding of the particle or the string, the role played by the group element $\Lambda$ is at this level quite obscure. In fact, we are about to see that this group variable encodes the momentum and spin of the matter excitation.

\section{Physical interpretation}

We now study the physical picture emerging from the types of matter theories considered above. 
We proceed in two steps. First, we discuss a limiting case of the above framework to establish a point of contact with the description of matter in terms of pseudo-classical variables, or Poincar\'e group coordinates in flat space \cite{Bal1}, \cite{Bal2}. 
Then, we study some examples showing how the formalism relates to the standart formulation of particles and strings on arbitrary backgrounds. 

\subsection{Flat limit}

In this section, we temporarily forget about the gravitational action and focus only on the matter sector of the theory.
To pinpoint the relation to the Poincar\'e or pseudo-classical description of matter, we will work in the zero cosmological constant limit of the above framework. In this limiting procedure, we have seen that the deSitter algebra $\so(\eta_5)$ contracts to the non-semi-simple Poincar\'e Lie algebra $i \so(\eta)$, and that the bilinear form $\langle , \rangle$ used to define the actions degenerates. Hence, we have to work with the bilinear form on the Poincar\'e algebra $(,)$ defined in \eqref{Poincare} which is only invariant under the Lorentz subgroup. So, the algebraic framework remains well defined\footnote{In fact, since the constant element $\U$ used to define the particle action lives in the dual of the Lie algebra, there is no need for any metric structure on the Lie algebra to define the action for a free particle in flat space. The free particle action can be entirely defined by using the dual pairing map. It is only the coupling to gravity which selects a particular bilinear form. This statement also holds for the string.} in the zero cosmological case.

The easiest way of understanding how the algebraic variables $(X,\Lambda)$ relate to ordinary notions such as position, momentum and spin is to fix a particular background geometry described by the Cartan connection $\mathbf{A}$. The simplest possible background is flat Minkowski spacetime which is obtained from the deSitter vacuum solution of the theory in the zero cosmological constant, i.e., infinite radius limit. We can now study the physics of the type of theories obtained above by evaluating the particle \eqref{particleMM}, and string \eqref{stringMM} actions on a such solution. 

Flat Minkowski spacetime corresponds to the connection 
\be
\A_{\mbox{{\tiny flat}}} = A \oplus e, \;\;\;\; \mbox{with} \;\;\;\; A = 0 \;\; \mbox{and} \;\; e = \delta_{\mu}^a dx^{\mu} \otimes P_a,
\ee
with the spacetime manifold $M$ identified with the internal space $V_{\eta}$.

When evaluating the matter actions \eqref{particleMM}, \eqref{stringMM} on the above background in the zero cosmological limit, both the particle and string actions share a common building block; the Poincar\'e valued one-form $\Omega_{X^* \mathbf{A}}$ evaluated on the flat connection pulled-back to the worldline or worldsheet
\be
\Omega_{X^* \mathbf{A}_{\mbox{{\tiny flat}}}}:=\Omega(X,\Lambda) = \Lambda^{-1} d \Lambda \oplus \Lambda^{-1} dX,
\ee
with $X$ and $\Lambda$ respectively describing the embedding of the particle or the string in flat space $V_{\eta}$ and in the isometry subgroup $\SO(\eta)$. But if $X$ and $\Lambda$ are regarded as coordinates on the Poincar\'e group manifold, this object can be related to a well known quantity; $\Omega$ is equal to the left-invariant Maurer-Cartan form $\Theta$ on the Poincar\'e algebra\footnote{The Poincar\'e Maurer-Cartan form $\Theta$ can be calculated by using the five by five matrix representation of the Poincar\'e group $U(\Lambda,X)= \left( \begin{array}{cc} \Lambda & X \\ 0 & 1 \end{array} \right)$, with $\Lambda$ in $\SO(\eta)$ and $X$ in $T^4 \simeq V_{\eta}$. The computation yields:
$$
\Theta = U^{-1} d U = \frac{1}{2}(\Lambda^{-1} d \Lambda)^{ab} \sigma_{ab} + (\Lambda^{-1} dX)^a P_a.
$$
}.
This is precisely the point of contact with the Poincar\'e or pseudo-classical description of matter, where the degrees of freedom of particles and strings are naturally parametrised by elements of the Poincar\'e group, or more precisely by a flat, pure gauge Poincar\'e connection $\Theta$ pulled-back to the worldline or worldsheet.

The fact that the degrees of freedom of massive, spinning matter in flat space are not entirely
captured by its position in space is an old idea going back, as far as we can see, to Finkelstein \cite{Finkelstein}. 
In this approach, the configuration space for a point particle or a string, or more generally
a $p$-brane, is the connected component of the Poincar\'e manifold $ISO(\eta)$. Hence, a $p$-brane 
is described by the embedding of it's worldsheet $U: \R^{p+1} \rightarrow ISO(\eta)$; $\tau \mapsto (X(\tau),\Lambda(\tau))$, in the Poincar\'e manifold. The translational part $X$ describes the position, 
while the Lorentz part $\Lambda$ encodes the spin and the momentum.
Now, when trying to construct an appropriate globally (left) Poincar\'e invariant and reparametrisation invariant Lagrangian for such degrees of freedom, a natural candidate appears to be the Maurer-Cartan form $\Theta$ on the Poincar\'e algebra because of its natural (left) Poincar\'e invariance. Considering the integral of the pull-back of the $p+1$ exterior power of this one-form to the worldsheet of the $p$-brane will naturally lead to a reparametrisation invariant and globally Poincar\'e invariant action. This is the point of contact between our formulation and the Poincar\'e description of matter.

To understand how this formalism indeed leads to the correct dynamics for particles and strings, we now compute the currents associated the actions \eqref{particleMM}, \eqref{stringMM} evaluated on a flat background.

\subsubsection{Particle}

In the $\Lambda \rightarrow 0$ limit, the evaluation of the particle action \eqref{particleMM} on the flat solution leads to the following globally Poincar\'e invariant and reparametrisation independent action 
\be
\label{particleflat}
S_{\mbox{{\tiny particle}}}[X,\Lambda] = \int_{\gamma} (\mathbf{u}, \, \Omega), 
\ee
where $\gamma$ is a curve in $I\SO(\eta)$, $\mathbf{u} = w \oplus v$, and $\Omega = \Lambda^{-1} d \Lambda \oplus \Lambda^{-1} dX$ are elements of $i\so(\eta) = \so(\eta) \oplus V_{\eta}$, and $( , )$ is the non-degenerate, $\SO(\eta)$-invariant symmetric bilinear form on $i \so(\eta)$ defined by \eqref{Poincare}.

When the Poincar\'e algebra element $\mathbf{u}$ is chosen conveniently, this action describes the dynamics of a massive, spinning point particle.
In fact, the above general action can describe many physical or unphysical particle dynamics depending on the value of $\mathbf{u}$, or more precisely, on the Lorentz subgroup adjoint orbit\footnote{More precisely, the physical content of the particle theory is determined by the Lorentz subgroup co-adjoint orbit in which the element $\U^* \in i\so(\eta)^*$ lies. However, since we are working with a bilinear form which is invariant under the adjoint action of the Lorentz subgroup, adjoint and coadjoint orbits under the Lorentz subgroup are naturally identified.} in which this element lies. By this we mean that different values of $\mathbf{u}$ living on the same Lorentz adjoint orbit will correspond to the same physical theory, because of the invariance of the bilinear form.
To understand these last points, we compute the currents associated to the theory defined by the action \eqref{particleflat}.

As a preliminary step, we remark that, using the identity 
$$\Lambda^{-1} d \Lambda = \Ad_{\Lambda^{-1}}(d \Lambda \Lambda^{-1}),$$
the action \eqref{particleflat} can be reexpressed as
\be
S_{\mbox{{\tiny particle}}}[X,\Lambda] = \int_{\gamma} ( \mathbf{J}, \, \omega ), 
\ee
with 
$$\omega = d \Lambda \Lambda^{-1}  \oplus dX \;\;\; \mbox{and} \;\;\; \mathbf{J}=\Ad_{\Lambda}(\mathbf{u}) = \Ad_{\Lambda}(w) \oplus \Lambda v,$$ 
by using the invariance of the bilinear form.

We are now ready to study the variational problem.
Varying the embedding $X$ of the world-line in coordinate space yields 
\bea
\delta S_{\mbox{{\tiny particle}}} &=& \int_{\gamma} ( \mathbf{J}, \, \delta \omega ) \nn \\
                                   &=& \int_{\gamma} ( \mathbf{J}, \, 0 \oplus d \delta X ) \nn \\
                                   &=& - \int_{\gamma} d ( \mathbf{J}, \, P_a ) \, \delta X^a,
\eea
where we have used the commutativity of the variation with the exterior differential $d \delta = \delta d$ in the second line, and an integration by parts 
in the last step.
Setting 
\be
p_a = -( \mathbf{J}, \, P_a ),
\ee 
the projection of the Poincar\'e algebra element $\mathbf{J}$ on the translational component,
the associated motion immediately implies a first conservation law 
\be
\delta S_{\mbox{{\tiny particle}}} = 0 \;\; \Rightarrow \;\; dp = 0,
\ee
that we naturally identify with momentum conservation.

Under an infinitesimal variation of the form $\delta \Lambda = \epsilon \circ \Lambda$, with $\epsilon$ an arbitrary element in $\so(\eta)$,
the Lorentz and translational component of the Maurer-Cartan form $\Omega$ respectively transform as 
\bea
\delta (\Lambda^{-1} d \Lambda) &=& \Ad_{\Lambda^{-1}}(d \epsilon) \nn \\
\delta (\Lambda^{-1} dX) &=& - \Lambda^{-1} \, \epsilon \, dX, \nn
\eea
and the action \eqref{particleflat} varies consequently
\bea
\delta S_{\mbox{{\tiny particle}}} 
&=& \int_{\gamma} ( \mathbf{J}, d\epsilon \oplus - \epsilon \, dX ) \nn \\
&=& \int_{\gamma} ( \mathbf{J}, d\epsilon)  + \, dX^a ( \mathbf{J}, P_b) \,  \epsilon^{ab} \nn \\
&=& - \frac{1}{2} \int_{\gamma} \left[ d( \mathbf{J}, \sigma_{ab}) + [dX \otimes p - p \otimes dX]_{ab} \right] \epsilon^{ab},
\eea
Here, we have used the Lorentz invariance of the bilinear form in the first step, and an integration by parts when going from the second to the third line.  

Setting 
\be
s_{ab} = \langle \mathbf{J}, \sigma_{ab} \rangle, 
\ee
and noting $dX \wedge p$ the element $[dX \otimes p - p \otimes dX]$ of $\Omega^2(V_{\eta})$,
the associated motion follows
\be
ds + dX \wedge p  = d (X \wedge p + s) = 0,
\ee
with the use of the first conservation law obtained above.
Hence, 
$$M = X \wedge p + s,$$ 
defines the second invariant of the particle which is therefore naturally identified with the total angular momentum, that is, the sum of the orbital and spin angular momenta. Hence, the variable $s$ carries the interpretation of the spin of the particle. 

To be able to conclude that the action \eqref{particleflat} describes the correct dynamics of a physical particle, we need to ensure that the variables $p$ and $s$ satisfy the appropriate requirements on momentum and spin. This is achieved by making the appropriate choice of the orbit in which the constant Poincar\'e element $\mathbf{u}$ lies. Indeed, a particle with positive mass $m$, timelike momentum and spin $\lambda$ corresponds to the following choice, up to Lorentz conjugation
\be
\mathbf{u} = \lambda \, \sigma_{12} \oplus -m P_0, 
\ee
with obvious generalisation to the massless and tachionic cases.
Rotating this element through the $\SO(\eta)$ action gives rise to the momentum and spin of the particle:
\bea
p &=& m \, \Lambda P_0 \;\; \in V_{\eta} \nn \\
s &=& \lambda \, \Ad_{\Lambda}(\sigma_{12}) \;\; \in  \, \so(\eta).
\eea
It is then immediate to show the $p$ and $s$ indeed satisfy, by construction, the physical requirements on the
momentum and spin of a massive, spinning particle with timelike momentum
\bea
p_a p^a &=& \sigma^2 \, m^2, \;\;\;\; p_0 > 0 \nn \\
p_a \, s^{ab} &=& 0 \nn \\
\frac{1}{2} s_{ab} s^{ab} &=& \lambda^2.
\eea

\subsubsection{String}

The flat limit of the string action \eqref{stringMM} yields the following globally Poincar\'e, and worldsheet reparametrisation invariant action 
\be
\label{stringactionflat}
S_{\mbox{{\tiny string}}}[X,\Lambda] = \int_{\Sigma} ( \mathbf{w}, \Omega \wedge \Omega ), 
\ee
where $\Sigma$ is now a surface embedded in $ISO(\eta)$, $( , ) := \frac{1}{2} ( , )^{\otimes 2}$, and $\W$ and $\Omega^{\wedge 2}$ are elements of $i \so(\eta) \otimes i \so(\eta)$.

As for the particle, the value of the Lorentz adjoint orbit in which the constant element $\mathbf{w}$ lies determines the physical content that the theory describes. To clarify this point, we determine the currents 
associated to the above action which will give rise to the momentum and spin of the string. These currents were computed in \cite{Stern}.

As we did above, we first derive a simplifying identity stating that the
the string action \eqref{stringactionflat} can be reexpressed as
\be
S_{\mbox{{\tiny string}}}[X,\Lambda] = \int_{\Sigma} ( \mathbf{K}, \omega \wedge \omega ), 
\ee
with 
$$\omega = d \Lambda \Lambda^{-1}  \oplus dX, \;\;\; \mbox{and} \;\;\;  \mathbf{K} = [\Ad_{\Lambda} \otimes \Ad_{\Lambda}](\mathbf{w}).$$
We are now ready to study the variational problem.

To this aim, we start by calculating the variation of the action with respect to variations of the embedding $X$ of the string
\bea
\label{variation}
\delta S_{\mbox{{\tiny string}}} &=& 2 \int_{\Sigma} \left( \mathbf{K} , \omega \wedge \delta \omega \right) \\
                                 &=& 2 \int_{\Sigma} \left( \mathbf{K} , 
\omega \wedge 
(0 \oplus \delta dX) \right) \nn \\
                                 &=& 2 \int_{\Sigma} d \left( \mathbf{K} , \omega
\otimes P_a \right) \, \delta X^a \nn.
\eea
Setting 
\be
\mathcal{P}_a = -( \mathbf{K}, \omega \otimes P_a ),
\ee
we have thus obtained the first invariant associated to the string by extremising the above variation
\be
\delta S_{\mbox{{\tiny string}}} = 0 \;\;\; \Rightarrow \;\;\; d \mathcal{P} = 0,
\ee
with the $V_{\eta}$-valued one-form $\mathcal{P}$ on $\Sigma$ consequently interpreted as the momentum density of the string.

Next, we compute the variation of the action under infinitesimal Lorentz variation of the form $\delta \Lambda = \epsilon \circ \Lambda$, with $\epsilon \in \so(\eta)$. Using the transformation properties established for the particle, it is straight-forward to obtain the variation of the action \eqref{stringactionflat}
\bea
\delta S_{\mbox{{\tiny string}}} &=& 2 \int_{\Sigma} \left( \W , \Omega \wedge \delta \Omega \right) \\ 
&=& 2 \int_{\Sigma} \left( \mathbf{K} , \omega \wedge (d\epsilon \oplus - \epsilon \, dX) \right) \nn \\
&=& \int_{\Sigma} \left[ d ( \mathbf{K}, \omega \otimes \sigma_{ab} ) 
- dX^a \wedge( \mathbf{K}, \omega \otimes P_b ) \right] \, \epsilon^{ab} \nn
\eea
Defining the $\so(\eta)$-valued one-form on $\Sigma$
\be
\s_{ab} = ( \mathbf{K}, \omega \otimes \sigma_{ab} ),
\ee
we obtain, extremising the above variation, the motion
\be
d\s + dX \wedge \p= d (X \wedge \p + \s)=0,
\ee
where the wedge product acts simultaneously on the cotangent bundle to the surface and on internal, or target space. 
This relation
expresses the conservation of the total angular momentum of the string 
$$\mathcal{M} = X \wedge \p + \s,$$
with the spin component given by $\s$.

Accordingly, the action \eqref{stringactionflat} describes the dynamics of a spinning string with momentum 
density $\p$ and spin $\s$, and the full action \eqref{stringMM} describes the interaction of a spinning string with gravity.

The situation is however not as clear as for the particle since the (squared) Poincar\'e algebra
element $\W$ is left unspecified, by lack of knowledge of the properties to be satisfied by strings admitting spin. 
We can nevertheless write down the value of $\W$ corresponding to the spinless, Nambu-Goto string. This theory corresponds to the choice \cite{Bal2}:
\be
\W = \frac{1}{2 \pi \alpha'} \, P_3 \wedge P_0 \;\;\;\; \in \;\; \Omega^2(V_{\eta}) \subset i \so(\eta)^{\otimes 2}, 
\ee
where $T=(1 / 2 \pi \alpha')$ is the string tension, $\alpha'$ being the Regge slope. 
With such a value of $\W$, computing the momentum and spin one-forms leads to 
\be
\mathcal{P}_a = \frac{1}{2 \pi \alpha'} [\Ad_{\Lambda}(P_3 \wedge P_0)]_{ab} \, dX^b \;\;\; \mbox{and} \;\;\; \mathcal{S}_{ab} = 0,
\ee
as expected for the Nambu-Goto string.
We will see in the next section how this theory is related to the Nambu string in further detail.

Having studied the flat space limit and clarified the interpretation of the variables $(X,\Lambda)$ extracted from the gauge symmetries of gravity, we now switch back on the gravitational interaction with positive cosmological constant.
The geometry is now fluctuating, no longer constrained to the flat solution, and switching on the cosmological constant has transformed the Poincar\'e algebra into the deSitter algebra. 
In this general case, we discuss the simplest possible examples with non-trivial background geometry.

\subsection{The spinless case}

The simplest examples of the types of theories considered here are the spinless particle and the Nambu-Goto string. We show how the corresponding actions relate to more standart formulations of particles and strings on arbitrary background geometries. We simply generalise to the MacDowell-Mansouri framework the flat space calculations done in \cite{Bal2}.

\subsubsection{Spinless particle}

A spinless particle coupled to MM gravity is described by an element $\mathbf{u}=0 \oplus v$ in $V \subset \so(\eta_5)$. In this case, the action \eqref{particleMM} reduces to
\be
\label{spinless}
S_{\mbox{{\tiny particle}}}[e;X,\Lambda] = \int_{\gamma} d \tau \, \mathcal{L}[e;X,\Lambda], \;\;\;\; \mbox{with} \;\; \mathcal{L}[e;X,\Lambda] = \langle \mathbf{J}, \, e_{\tau} \rangle,
\ee
where $e_{\tau} :=  e_{\mu}^a P_a \partial_{\tau} X^{\mu}$ is the image in $V$ of the (timelike) tangent vector to the curve, and $\mathbf{J} = \Ad_{\Lambda}(\mathbf{u}) = -l^2 p$ coincides with the momentum $p$ of the particle, rescalled with the cosmological length to give the action the correct dimension.

The adjoint orbit of the Lorentz subgroup in which $\mathbf{u}$, or equivalently $\mathbf{J}$, lies is labelled by the invariant
$$
C = \langle \mathbf{J}, \mathbf{J} \rangle,
$$
and determines what kind of spinless particle the theory describes, i.e, wether the particle is massive, massless or tachyonic. A massive particle with mass $m$ and timelike momentum corresponds to the orbit labelled by
\be
C = \sigma^2 \, l^2 m^2.
\ee 
The constant element $\mathbf{u}:=\mathbf{u}(m,0)=m l^2 P_0$ is then a chosen representative of the associated conjugacy class and characterises the theory. 

Now let us show how this choice indeed leads to the correct action of a massive particle.
Varying $\Lambda$ as we did in the flat case leads to $\delta \mathbf{J} = [\epsilon, \mathbf{J}]$, and the associated variation follows
\bea
\delta \mathcal{L} &=& \langle [\epsilon, \mathbf{J}], e_{\tau} \rangle \nn \\
                                   &=& \langle \epsilon, [\mathbf{J}, e_{\tau}] \rangle,
\eea
where we have used the full deSitter invariance of the bilinear form.
The corresponding motion yields
$$
- \frac{1}{l^2} \, \mathbf{J} \wedge e_{\tau} = 0,
$$
with the wedge product acting on internal space.
This equation states that $\mathbf{J}$, or equivalently the momentum $p$, and $e_{\tau}$ are parallel (in internal space).

Thus, there exists an field $\alpha$ 
such that $\mathbf{J} = \alpha \, e_{\tau}$. We can solve for $\alpha$ by exploiting the mass shell constraint 
$$
\langle \mathbf{J}, \mathbf{J} \rangle = \alpha^2 \langle e_{\tau}, e_{\tau} \rangle = \sigma^2 l^2 m^2.
$$
Hence, one is led to the following values of $\mathbf{J}$:
\be
\mathbf{J} = \alpha \, e_{\tau}, \;\;\;\; \mbox{with} \;\; \alpha = \pm \frac{m l}{\sqrt{\sigma^2 \langle e_{\tau}, e_{\tau} \rangle}}.
\ee
To solve the sign ambiguity, one has to supplement the particle action with the sign constraint $\sigma^2 \mathcal{L} > 0$ on the Lagrangian, ensuring that if the worldline is future pointing, the particle will have positive energy.

Then, the positive solution is selected, and reinserting $\mathbf{J}$ into the original action \eqref{spinless} leads to the standart action for a massive spinless particle
\be
S_{\mbox{{\tiny particle}}}[e;X] = \sigma^2 m \, L[\gamma,e] = \sigma^2 m \int_{\gamma} d \tau \sqrt{\sigma^2 \eta(e_{\tau},e_{\tau})},
\ee
that is, the length $L[\gamma,e]$ of the worldline $\gamma$, measured in the geometry determined by the co-frame $e$, weighted by the mass $m$ of the particle.

\subsubsection{Nambu-Goto string}

Similarly, we can obtain the Nambu-Goto string through an appropriate adjoint orbit choice.
A spinless string is described by an element $\mathbf{w} = v_1 \otimes v_2$ whose antisymmetric part lies in $\Omega^2(V)$, with $\Omega^2(V)$ included in $\so(\eta_5) \otimes \so(\eta_5)$ as a vector subspace.

Choosing a local basis  $\{\partial / \partial \tau^{\alpha} \}_{\alpha=0,1}$ of the tangent space $T_p \Sigma$, with $p$ in $\Sigma$, the corresponding action yields
\be
\label{spinlesstring}
S_{\mbox{{\tiny string}}}[e,X,\Lambda] = \int_{\Sigma} d^2 \tau \, \mathcal{L}, \;\;\;\; \mbox{with} \;\; \mathcal{L} = \langle \mathbf{K}, B \rangle,
\ee
where 
$$
B := \, e^a_{\mu} e^b_{\nu} P_a \wedge P_b \, \Sigma^{\mu \nu}, \;\;\;\; \Sigma^{\mu \nu}=\epsilon^{\alpha \beta} \partial_{\alpha} X^{\mu} \partial_{\beta} X^{\nu},
$$ 
is the image in inertial space of the bivector $\Sigma^{\mu \nu}$ associated to the surface $\Sigma$, and $\mathbf{K} = [\Ad_{\Lambda} \otimes \Ad_{\Lambda}](\mathbf{w})$. The bivector $\Sigma^{\mu \nu}$ is assumed to be constructed from a timelike and a spacelike vector.
As for the particle, we supplement the string Lagrangian with the strict positivity constraint $\sigma^2 \mathcal{L} > 0$.

The type of string theory defined by the action \eqref{spinlesstring} depends on the Lorentz subgroup orbit in the space of (internal) bivectors $\Omega^2(V)$ in which $\mathbf{K}$, or equivalently $\mathbf{w}$ lies. As already remarked, the space $\Omega^2(V)$, being a real six dimensional vector space, is naturally identified with the isometry algebra $\so(\eta)$ and carries the adjoint representation. Furthermore, the bilinear form $\langle , \rangle$ on $\so(\eta_5) \otimes \so(\eta_5)$ reduces on $\Omega^2(V)$ to the isometry algebra Killing form, up to rescalling, and can thus be used to characterise the orbit space of the adjoint action of the isometry subgroup $\SO(\eta)$ on $\Omega^2(V)$. 
The orbit in which $\mathbf{K}$ lies is labelled by the two invariants
$$
C_1= \langle \mathbf{K} , \mathbf{K} \rangle \,, \;\;\; \mbox{and} \;\;\; C_2 = \langle \, * \mathbf{K} , \mathbf{K} \rangle ,
$$
related to the quadratic and pseudo-scalar Casimirs of the isometry algebra $\so(\eta)$.
A choice of orbit specifies if the string is the Nambu-Goto, tachyonic or null string \cite{Bal2}.
As anticipated above, the orbit corresponding to the standart Nambu-Goto string is the following
\be
\label{casimirs}
C_1 = \sigma^2 (\frac{l^2}{2 \pi \alpha'})^2, \;\;\; \mbox{and} \;\;\; C_2=0.
\ee
The constant element $\mathbf{w}=(l^4 / 2 \pi \alpha') \, P_3 \wedge P_0$ is then a chosen representative of the associated conjugacy class and thus determines the theory.

To show that the above choice of $\mathbf{w}$ leads to the Nambu-Goto action, we proceed to the variation of the action \eqref{spinlesstring} with respect to the field $\Lambda$ as we did for the particle. A general variation of $\Lambda$ leads to the variation $\delta \mathbf{K} = [\epsilon, \mathbf{K}]$ of $\mathbf{K}$, and the associated motion follows
$$
\left[ \, \mathbf{K} , B \, \right] = 0 \;\; \Leftrightarrow \;\; \mathbf{K} \in C(B).
$$
If $\mathbf{K}$ is regarded as an element of $\so(\eta)$, solving this equation for $\mathbf{K}$ is equivalent to finding the centraliser $C(B)$ of $B$ in $\so(\eta)$. Using the fact that $[X, * X] = * [X,X]=0$ forall $X$ in $\so(\eta)$, one can show that $C(B)$ is of dimensions two (this is a reflection of the rank of the isometry algebra) and generated by $B$ and $*B$ as a real vector space. Hence, $\mathbf{K}$ is necessarily of the form
\be
\mathbf{K} = \beta B + \gamma * B.
\ee
This condition is the analogue of the parallelism of the momentum and (image in internal space of the) velocity vectors obtained earlier for the point particle.
Since the B field is {\em simple}, that is, of the form $e \wedge e$, it automatically satisfies $\langle * B, B \rangle= 0$.
Hence, from the condition \eqref{casimirs} on the invariants, one is led to
$$
(\beta^2 + \sigma^2 \gamma^2) \langle B , B \rangle  = \sigma^2 (\frac{l^2}{2 \pi \alpha'})^2, \;\;\; \mbox{and} \;\;\; 2 \sigma^2 \beta \, \gamma \langle B , B \rangle = 0,
$$
where we have used $*^2 = \sigma^2$, and $\langle * B , * B \rangle = \sigma^2 \langle B , B \rangle$.
The first equation tells us that $\langle B , B \rangle$ is different from zero. Thus, from the second we obtain that either $\beta$ or $\gamma$ is zero. If $\beta=0$, plugging the value of $\mathbf{K}$ into the action yields a vanishing action by simplicity of the B field, which contradicts the strict positivity requirement on the Lagrangian. Thus, $\gamma=0$ and we are left with 
\be
\mathbf{K} = \beta B, \;\;\;\; \mbox{with} \;\; \beta = \pm \frac{l^2}{2 \pi \alpha'} \frac{1}{\sqrt{\sigma^2 \langle B , B \rangle}}.
\ee
The sign constraint on the Lagrangian implies that only the positive solution is admissible. Then, 
inserting the obtained value of $\mathbf{K}$ in the action leads to the Nambu-Goto action for string theory on 
a curved background
\be
S_{\mbox{{\tiny string}}}[e,X] = \sigma^2 \frac{1}{2 \pi \alpha'} A[\Sigma,e] ,
\ee
where 
\be
A[\Sigma,e]=\int_{\Sigma} \, d^2 \tau \, \sqrt{-\frac{\sigma^2}{2} \tr B^2},
\ee
is the area of the surface $\Sigma$ measured with respect to the background geometry determined by the tetrad $e$.
This is because the trace of $B^2$ is related to the metric induced on the surface by
$$
-\frac{1}{2} \tr B^2 = \det X^* g,
$$
with $g = \eta (e \otimes e)$ the metric on $M$.

To close this discussion on the physical interpretation of the formalism, we summarise the main lessons.
The actions \eqref{particleMM}, \eqref{stringMM} describe the coupling of general particles and strings to gravity.
The matter variables $X$ and $\Lambda$, extracted from the symmetries of the gravitational interaction, respectively encode the embedding and the momentum and spin of the particle or string. The value of the latter physical quantities is dictated by the classifier elements $\U$ or $\W$. For appropriate values, the theory describes massive particles with spin or the Nambu-Goto string augmented with spinning contributions.

Finally, it is important to stress that the above derivation linking the spinless framework to more conventional formulations of particles and strings is not tied to the presence of a cosmological constant and to the Mac-Dowell-Mansouri formulation.
In the zero cosmological constant case, one simply needs to replace the bilinear forms in the calculations presented above with their Poincar\'e analogues, which will only involve the metric on the vector representation and the trace on the Lorentz algebra in the particle and string case respectively. The results obtained will be the same than the ones obtained in the deSitter case by setting $l=1$ in the choice of the orbits and in the calculations that follow.

\section{Variational problem of the coupled system}

We are now ready to study the motions of the full theory defined by the GR action \eqref{MM} augmented with, firstly, the particle action \eqref{particleMM} and secondly the string action \eqref{stringaction}. We start by considering the matter sector before computing the variations of the gravitational variables.

\subsection{Matter motion}

Based on the flat space experience, we now
firstly compute the variations of the action \eqref{particleMM} describing a particle coupled to gravity. Then, we will consider the variations of the string action \eqref{stringMM}.

\subsubsection{Particle}

Under an infinitesimal variation of the Lorentz group variable $\delta \Lambda$, the gauged Maurer-Cartan form $\Omega_{\A}$ transforms as $\delta \Omega_{\A} = \Ad_{\Lambda^{-1}}(d_{\mathbf{A}} \epsilon)$, and the variation of the action \eqref{particleMM} consequently yields
\bea
\delta S_{\mbox{{\tiny particle}}} &=& \int_{X(\gamma)} \langle \mathbf{u}, \delta \Omega_{\A} \rangle \nn \\
&=& \int_{X(\gamma)} \langle \mathbf{J}, d_{\mathbf{A}} \epsilon \rangle \nn \\
&=& - \frac{1}{2} \int_{\gamma} \langle \nabla \mathbf{J}, \sigma_{ab} \rangle \, \epsilon^{ab},
\eea
where $\nabla \equiv d_{X^* \mathbf{A}}$ is the covariant derivative of the deSitter connection pulled back to the worldline, 
and we have used the full deSitter invariance of the bilinear form in the last step.

Using the results on the flat space analysis and the spinless example, we know that the deSitter element $\mathbf{J}$ is the sum of the momentum and spin of the particle: 
\be
\mathbf{J} = s \oplus -l^2 p.
\ee
We will need to compute the covariant derivative of this current. For clarity purposes, we write the corresponding expression, and many of the following, in local, component language:
$$
\nabla \mathbf{J} = \left( \frac{1}{2} \left[ (D_{\tau} s)^{ab} + 2 \, e_{\tau}^{[a} \, p^{b]} \, \right] \sigma_{ab} -\left[ l^2 (D_{\tau} p)^a + s^{ab}  e_{\tau b} \right] P_a \right) d \tau,
$$
where 
$$
D_{\tau} s^{ab}= \partial_{\tau} s^{ab} + [A_{\tau},s]^{ab}, \;\;\; \mbox{and} \;\;\; D_{\tau} p^{a}= \partial_{\tau} p^{a} + A^{a}_{\tau \; b} \, p^b.
$$
We have introduced the notation $X^* \eta := \eta_{\tau} d \tau$, with $\eta_{\tau} = \eta_{\mu} \partial_{\tau} X^{\mu}$ forall one-form $\eta$ on $M$.

It is now immediate to obtain the first set of equations of motion of a massive spinning particle coupled to gravity 
\be
\label{precession}
\delta S_{\mbox{{\tiny particle}}} = 0 \;\;\;\;\; \Leftrightarrow \;\;\;\;\; 2 \, e_{\tau}^{[a} \, p^{b]} +  D_{\tau} s^{ab} = 0,
\ee
known as the spin precession equation.

Note that this equation is simply the curved space generalisation of the flat space conservation of total angular momentum obtained earlier, in the the zero cosmological constant case. 

Next, we consider variations of the embedding $X$ of the worldline $\gamma$, that is, the variation under an infinitesimal diffeomorphism bending the curve $\gamma$. As in the flat case, it is convenient to define the conjugate variable $\omega_{\A} = \Ad_{\Lambda}(\Omega_{\A})$. The variation of the one-form $\omega_{X^* \A}$
under an infinitesimal variation of $X$ yields 
$$
\delta \omega_{X^* \A} = \delta (X^* \mathbf{A}) = X^* \mathcal{L}_{\delta X} \mathbf{A},
$$ 
with $\mathcal{L}_{\delta X} = i_{\delta X} d + d i_{\delta X}$ the Lie derivative along the vector field $\delta X$, 
with $i$ here denoting the interior product on $\Omega(M)$. Hence, 
\bea
\label{papapetrou}
\delta S_{\mbox{{\tiny particle}}} 
&=& \int_{\gamma} \langle \mathbf{J}, X^* \mathcal{L}_{\delta X} \mathbf{A} \rangle \\
&=& \int_{\gamma} \langle \mathbf{J}, X^* \left[ d_{\A} (i_{\delta X}(\A)) + i_{\delta X}(\F) \right] \rangle \nn \\
&=& \int_{\gamma} - \langle \nabla \mathbf{J}, i_{\delta X}(\A) \rangle + \langle \mathbf{J}, X^* i_{\delta X}(\F) \rangle, \nn
\eea
where we have used the equivalence between infinitesimal gauge transformations, with field dependent parameter $\alpha=i_{\delta X}(\A)$, and infinitesimal diffeomorphisms up to the curvature of the connection 
\be
\label{lie}
\mathcal{L}_{\delta X} \A = d_{\A} \alpha + i_{\delta X}(\F),
\ee
in the third line. Using the equations on spin precession, the above equation of motion can be reduced to the Mathison-Papapetrou equations \cite{Mathison} with torsion
\be
D_{\tau} \, p^a \, e_{\mu \, a} + ( \frac{1}{2} s^{ab} F_{\mu \nu \, ab} - p^a T_{\mu \nu \, a}) \partial_{\tau} X^{\nu} = 0,
\ee
where $T = d_A e$ is the torsion two-form, and $F \equiv F_A$ is the curvature of the Lorentz connection.
See \cite{Mathison}, \cite{Bal1}, \cite{artem} for details.

\subsubsection{String}

As for the particle, we start by studying variations of the action under infinitesimal variations of the group variable $\Lambda$. The variation of the action \eqref{stringMM} yields
\bea
\label{motion}
\delta S_{\mbox{{\tiny string}}} &=& 2 \int_{X(\Sigma)} \langle \mathbf{w}, \Omega_{\mathbf{A}} \wedge \delta \Omega_{\mathbf{A}} \rangle \\
&=& 2 \int_{X(\Sigma)} \langle \mathbf{K}, \omega_{\A} \wedge d_{\mathbf{A}} \epsilon \rangle \nn \\
&=& \int_{X(\Sigma)} \left[ d \langle \mathbf{K}, \omega_{\mathbf{A}} \otimes \sigma_{ab} \rangle 
+ \langle \mathbf{K}, \omega_{\mathbf{A}} \wedge [\mathbf{A} , \sigma_{ab}] \rangle \right] \epsilon^{ab}, \nn
\eea
where $\omega_{\A}=d_{\mathbf{A}} \Lambda \, \Lambda^{-1}$. 

We define the following pair of one-forms on $\Sigma$
\be
\boldsymbol{S}_{ab} = \langle \mathbf{K}, \omega_{X^* \mathbf{A}} \otimes \sigma_{ab} \rangle, \;\;\; \mbox{and} \;\;\;
\boldsymbol{P}_{a} = -\langle \mathbf{K}, \omega_{X^* \mathbf{A}} \otimes P_{a} \rangle,
\ee
respectively interpreted as the curved spacetime\footnote{Note that, unlike the particle, the momentum and spin of the string depend on derivatives of the variables $X$ and $\Lambda$. Thus, through a minimal coupling procedure, these variables depend on the gravitational field and their flat values are thus modified on curved backgrounds. The other way round, evaluated on the flat solution, the curved momentum and spin $\boldsymbol{P}$ and $\boldsymbol{S}$ yield their flat values $\mathcal{P}$ and $\mathcal{S}$.} value of the spin and momentum of the string.
We will use the local expression of their covariant derivative in the Lorentz connection $A$
$$
D_{\alpha} \boldsymbol{S}_{\beta}^{\, ab} = \partial_{\alpha} \boldsymbol{S}_{\beta}^{\, ab} + [A_{\alpha}, \boldsymbol{S}_{\beta}]^{ab}, \;\;\; \mbox{and} \;\;\; D_{\alpha} \boldsymbol{P}_{\beta}^{\, a} = \partial_{\alpha} \boldsymbol{P}_{\beta}^{\, a} + A^{a}_{\alpha \; b} \boldsymbol{P}_{\beta}^{\, b},
$$
where we have introduced the notation $X^* \eta = \eta_{\alpha} d \tau^{\alpha}$, with
$\eta_{\alpha} = \eta_{\mu} \partial_{\alpha} X^{\mu}$, forall one-form $\eta$ on $M$.

Now, extremising the variation derived above \eqref{motion}, we obtain a spin precession equation for the string. Written in local coordinates, the motion is the following
\be
\epsilon^{\alpha \beta} \left[ 2 \, e_{\alpha}^{[a} \boldsymbol{P}_{\beta}^{\, b]} + D_{\alpha} \boldsymbol{S}_{\beta}^{\, ab} \right]= 0.
\ee
Finally, we study the motion obtained by varying the embedding of the string.
Using the identities derived for the particle, it is immediate to compute the variation of the pull-back to the surface of the two-form $\omega_{\A}^{\wedge 2}$,
and the variation of the action follows
\bea
\delta S_{\mbox{{\tiny string}}} &=& \int_{\Sigma} \langle \mathbf{K}, \delta (X^*(\omega_{\A} \wedge \omega_{\A})) \rangle \\
&=& 2 \int_{\Sigma} \langle \mathbf{K}, X^*(\omega_{\A} \wedge \mathcal{L}_{\delta X} \mathbf{A}) \rangle \nn \\
&=& \int_{\Sigma} \left[ d \langle \mathbf{K}, \omega_{X^* \A} \otimes \sigma_{AB} \rangle + \langle \mathbf{K},  X^* (\omega_{\A} \wedge [\A, \sigma_{AB}]) \rangle \right] i_{\delta X}(\A^{AB}) \nn \\
&+& 2 \int_{\Sigma} \langle \mathbf{K}, X^*(\omega_{\A} \wedge i_{\delta X}(\F)) \rangle, \nn
\eea
where we have used the relation \eqref{lie} between the Lie derivative and the spacetime curvature, together with a partial integration in the last step.

Written locally, the motion associated to the above variation yields
\bea
0 &=& \int_{\Sigma} \, d^2 \tau \epsilon^{\alpha \beta} \delta X^{\mu} \left[  \left( D_{\alpha} \boldsymbol{S}_{\beta}^{\, ab} + 2\, e_{\alpha}^{[a} \boldsymbol{P}_{\beta}^{\, b]} \right) A_{\mu \, ab} \right. \nn \\
&-& 2 \, D_{\alpha} \boldsymbol{P}_{\beta}^{\, a} \, e_{\mu \, a} + \frac{1}{l^2} \boldsymbol{S}_{\alpha}^{\, ab} \, (e_{\mu \, a} \, e_{\nu \, b} - e_{\mu \, b} \, e_{\nu \, a}) \, \partial_{\beta} X^{\nu} \nn \\
&+& \left. \left( \boldsymbol{S}_{\alpha}^{\, ab} (F_{\mu \nu \, ab} - \frac{1}{l^2} (e_{\mu \, a} e_{\nu \, b} - e_{\mu \, b} e_{\nu \, a})) - 2 \boldsymbol{P}_{\alpha}^{\, a} T_{\mu \nu \, a} \right) \partial_{\beta} X^{\nu} \right]. \nn
\eea
Using the spin precession equation, we are naturally led to a Mathison-Papapetrou equation with torsion for the string
\be
\epsilon^{\alpha \beta} \left[ D_{\alpha} \boldsymbol{P}_{\beta}^{\, a} \, e_{\mu \, a} - (\frac{1}{2} \boldsymbol{S}^{\, ab}_{\alpha} F_{\mu \nu \, ab}  - \boldsymbol{P}^{\, a}_{\alpha} T_{\mu \nu \, a} ) \partial_{\beta} X^{\nu} \right] = 0.
\ee
Note that in the flat limit, this equations simply reduces to the conservation of the momentum density $\mathcal{P}$.

\subsection{Gravitational field equations}

To finish the study of the classical aspects of the theories introduced in this paper, we now calculate the gravitational motion defining the equations determining the gravitational field created by point and string sources. 

The variation of the gravitational sector yields
\bea
\delta S_{\mbox{{\tiny MM}}} &=& 2 \alpha \int_M \langle * \phi[F_{\mathbf{A}}] \wedge \delta \phi[F_{\mathbf{A}}] \rangle \\
&=& 2 \alpha \int_M \langle * (F_{A} - \frac{\Lambda}{3}  e \wedge e) \wedge ( d_A \delta A - \frac{2 \Lambda}{3}  e \wedge \delta e ) \rangle \nn \\
&=& - \frac{1}{4 \pi G} \int_M \langle * (e \wedge d_A e) \wedge \delta A \rangle 
+ \langle * (F_{A} - \frac{\Lambda}{3}  e \wedge e) \wedge e \wedge \delta e \rangle. \nn 
\eea
We have used an integration by parts together with the following expression for
the covariant derivative of the Lorentzian part of the deSitter curvature: 
$$
d_A (F_{A} - \frac{\Lambda}{3}  e \wedge e) = d_A F_A - \frac{\Lambda}{3} (d_A e \wedge e - e \wedge d_A e) = \frac{2 \Lambda}{3} e \wedge d_A e.
$$
Next, we introduce the matter sources described by the particle and string action.

We start by considering the back-reaction of a particle on the gravitational field which is described by the McDowell-Mansouri action \eqref{MM} augmented with the particle term \eqref{particleMM}.
It is more convenient to work with the action written in terms of the dual variable $\omega_{\A}$, the variation of which we now compute.
\bea
\delta S_{\mbox{{\tiny particle}}} &=& \int_{X(\gamma)} \langle \mathbf{J}, \delta \omega_{\A} \rangle \\
&=& \int_{X(\gamma)} \langle \mathbf{J}, \delta A \oplus \delta e \rangle \nn \\
&=& \int_{X(\gamma)} -\frac{1}{2} \tr s \, \delta A - \eta(p , \delta e ). \nn 
\eea
It is now immediate to write the corresponding back-reaction on the spacetime geometry
\bea
\epsilon^{\mu \nu \rho \sigma} \epsilon_{abcd} \, e^c_{\nu} \, T_{\rho \sigma}^d &=& - 8 \pi G \int_{\gamma} s_{ab} \, d X^{\mu} \, \delta(x - X(\tau)) \\
\epsilon^{\mu \nu \rho \sigma} \epsilon_{abcd} \, e_{\nu}^b ( F_{\rho \sigma}^{cd} - \frac{2 \Lambda}{3} e_{\rho}^{[c} \, e_{\sigma}^{d]}) &=& - 16 \pi G \int_{\gamma} p_{a} \, d X^{\mu} \, \delta(x - X(\tau)) \nn ,
\eea
where each field on the left hand side is evaluated at the point $x \in M$. 
These equations, if solvable, characterise the gravitational field produced by a particle of momentum $p$ and spin $s$.
However, to our knowledge, there are no known solutions to these equations. 

On the other hand, we can proceed in analogous way to calculate the effect of strings on the spacetime geometry. Varying the string action \eqref{stringMM}, we obtain
\bea
\delta S_{\mbox{{\tiny string}}} &=& 2 \int_{X(\Sigma)} \langle \mathbf{K}, \omega_{\A} \wedge \delta \omega_{\A} \rangle \\
&=& 2 \int_{X(\Sigma)} \langle \mathbf{K}, \omega_{\A} \wedge ( \delta A \oplus \delta e ) \rangle \nn \\
&=& 2 \int_{X(\Sigma)} -\frac{1}{2} \tr \boldsymbol{S} \wedge \delta A  - \eta(\boldsymbol{P} \wedge \delta e), \nn 
\eea
with the wedge product acting on the co-tangent bundle to the surface.
Thus, we can calculate the gravitational field produced by a string of momentum density $\p$ and spin $\s$. It is defined by the Einstein equations
\bea
\epsilon^{\mu \nu \rho \sigma} \epsilon_{abcd} \, e^c_{\nu} \, T_{\rho \sigma}^d &=& - 16 \pi G \int_{\Sigma} \boldsymbol{S}_{ab} \wedge d X^{\mu} \, \delta(x - X(\tau)) \\
\epsilon^{\mu \nu \rho \sigma} \epsilon_{abcd} \, e_{\nu}^b ( F_{\rho \sigma}^{cd} - \frac{2 \Lambda}{3} e_{\rho}^{[c} \, e_{\sigma}^{d]}) &=& - 32 \pi G \int_{\Sigma} \boldsymbol{P}_{a} \wedge d X^{\mu} \, \delta(x - X(\tau)) \nn .
\eea
Unlike the point particle, it is possible to find explicit solutions to the above equations, at least in the zero cosmological constant limit and for some specific values of the deSitter element $\mathbf{w}$. These solutions are called cosmic strings.

A cosmic string is the lift to $3+1$ dimensions of the point particle solution in $2+1$ 
gravity. It corresponds to a spacetime around an infinitely thin and long straight string 
(see for instance \cite{deser} and references therein).
In an earlier paper \cite{AW}, we showed that in the
spinless and zero cosmological constant case, an action leading to the motion solved by the 
cosmic string could be written. This action is precisely of the form of the spinless action
\eqref{spinlesstring} derived as a particular case of the general framework developed here, but in
the BF gravity context (see next section).
We will not reiterate the analysis and refer the interested reader to reference \cite{AW} 
where the statement is shown in detail.
What is new is the link with the Nambu-Goto string. There is a
free parameter, the mass per unit length $\mu$ of the string, entering the equations describing the 
gravitational field around a cosmic string. To obtain a match with the equations obtained above,
the classifier element $\mathbf{w}$ must be tuned appropriately. It is
interesting to remark that the cosmic string solution tunes $\mathbf{w}$ to be proportional to
$\mu P_3 \wedge P_0$, which is precisely the value describing the Nambu-Goto string, if we identify
the mass per unit length with the tension of the string. This shows a
link between the Nambu-Goto string and cosmic string solutions of general relativity.
One would need to see if these results extend to the spinning, non-zero cosmological constant case.
We leave these investigations for further work. 

To conclude the core of the paper, we explain how to relate the framework developed here to pioneering approaches describing the coupling of strings to BF-like theories. This leads us to discuss the perspectives offered by such a formalism.

\section{Strings in BF-type theories - Perspectives}

Gauge defects along two-surfaces were firstly considered \cite{BFloops}, \cite{BP} in the context of BF theory,
a topological field theory of Schwartz type.
The idea was to exploit the two-form nature of one of the fundamental fields of the theory to introduce matter excitations along surfaces. Allthough interesting, the obtained models clearly lacked a physical interpretation; the matter sources were string-like but did not seem to be related to any known formulation of string theory. An important conceptual progress was achieved in \cite{AW}, but the link with string theory structures was still not understood. One of the original motivations for this work was to explore if such a correlation existed.

So, what does the framework presented in this paper teach us about string-like sources coupled to BF theory as described in \cite{BFloops}, \cite{BP}, \cite{AW}, \cite{Ale}?
Well, retrospectively, it now appears clearly that such surface defects are related to the Nambu-Goto string once that some constraints are implemented in BF theory. These constraints introduce local degrees of freedom in the theory and transform BF theory into general relativity.
 
BF theory is a topological field theory of Schwartz type whose fundamental fields in four dimensions are a two-form $B$ valued in the Lie algebra $\mathfrak{g}$ of a chosen semi-simple Lie group $G$ and a $G$-connection $A$. In the vanishing cosmological constant case, the action is
\be
S_{\mbox{{\tiny BF}}}[B,A] = \frac{1}{\kappa} \int_{M} B \wedge F_{A} , 
\ee
where the algebra indices are paired with an $\Ad$-invariant bilinear form on $\mathfrak{g}$, and $\kappa \in \R$ is a coupling constant.
Extended matter can be naturally coupled to BF theory by considering gauge defects along surfaces of the form $\int_{\Sigma} \tr v B$, where $v$ is a fixed vector in $\mathfrak{g}$, before performing an inverse gauge transformation on the $B$ field. This leads to the action considered in \cite{BP}, \cite{AW}, \cite{Ale}. Now the resulting theory can not be interpreted as a string theory because of its topological character. But there is a way of introducing local degrees of freedom in BF theory. Then, BF theory becomes gravity and the string source becomes the Nambu-Goto string. Let us discuss this point in further details.

By comparing the Einstein-Cartan action \eqref{pureaction} with vanishing cosmological constant to the above BF action with structure group $G=\SO(\eta)$ and bilinear form $-\frac{1}{2} \tr$, it is easy to see that the two actions simply match provided the $\kappa = 8 \pi G$ and the following identification holds
$$
B = * e \wedge e.
$$
The remarkable fact, discovered by Plebanski \cite{Pleb}, is that this constraint can be imposed in the BF theory action. Let us simply pinpoint the result (see for instance \cite{FP} for details). The idea is to add to the BF action a quadratic term $\mathcal{C}[\Psi,B]$ in the B field with internal indices contracted with a Lagrange multiplier zero-form $\Psi$ satisfying appropriate symmetry requirements\footnote{Explicitly, the constraint term is of the form $\mathcal{C}[\Psi,B] = -\frac{1}{4} \int_{M} \Psi_{abcd} B^{ab} \wedge B^{cd}$, with $\Psi$ taking value in the symmetric part of $\so(\eta) \otimes \so(\eta)$, i.e., $\Psi_{[abcd]} = 0$.}. Varying the Lagrange multiplier $\Psi$ in the action
\be 
\label{plebanski}
S_{\mbox{{\tiny Plebanski}}}[B,A,\Psi] = \frac{1}{8 \pi G} S_{\mbox{{\tiny BF}}} + \mathcal{C}[\Psi,B],
\ee
forces $B$ to be simple, i.e., in one of the four sectors 
$$
B = \pm e \wedge e \;\;\;\; (I_{\pm}), \;\;\; \mbox{or} \;\;\; B = \pm * e \wedge e \;\;\;\; (II_{\pm}).
$$
The first set of B field configurations corresponds to a topological sector, while the second set indeed describes gravity. So, first order general relativity corresponds to one of the sectors of the theory derived from the action \eqref{plebanski}.

Now, using the methods presented in section $3.2.2$, one can show that an action of the form 
$$
S[B; X, \Lambda] = - \frac{k}{4 \pi \alpha'} \int_{X(\Sigma)} \tr \Ad_{\Lambda}(w) B,
$$
with $w = \sigma_{30}$, and $k=\pm 1$ (resp. $k=\pm \sigma^2$) when the $B$ field is in the $I_{\pm}$ sector (resp. in the $II_{\pm}$ sector), reproduces the Nambu-Goto action when varying $\Lambda$ and reintroducing the solution of the motion back in the action. Hence, plugging any sector of $B$ field solutions to the Plebanski constraints in the above action provides a first order formulation of the Nambu-Goto string.

Accordingly, the action obtained by adding the above source term to the Plebanski action 
\bea
S[B,A,\Psi;X,\Lambda] &=& - \frac{1}{16 \pi G}\int_M \tr \, B \wedge F_{A} + \mathcal{C}[\Psi,B] \\
&-& \frac{k}{4 \pi \alpha'} \int_{X(\Sigma)} \tr \Ad_{\Lambda}(w) B \nn,
\eea
with appropriate choice of the constant $k$, describes the Nambu-Goto string coupled either to 4D gravity, or to a topological theory. To suppress the topological sector, one should extend the framework to introduce an Immirzi parameter in the theory. We postpone this generalisation to future work.

Now the key point is that this BF formulation of strings coupled to 4D gravity appears to be promising from the spinfoam quantum gravity perspective.

Spinfoam models \cite{SF} define a regularisation of the path integral of gravity based on the convergence of different fields of research, such as topological field theory, simplicial quantum gravity, lattice gauge theory and loop quantum gravity. Now, all known spinfoam models of quantum gravity are based on the Plebanski formulation of GR as a constrained topological theory.
Hence, the framework presented here provides a spinfoam-compatible formulation of (spinless) strings, in the sense that it only involves structures that are explicitly exportable to the spinfoam framework. Furthermore, in spinfoam models, spacetime curvature is concentrated along two-surfaces, which implies that strings provide the most natural candidates for curvature defects.
We are currently investigating the construction of such models \cite{ARW}. 

\section{Conclusion} 

In this paper, we have studied the inclusion of point and string matter in four-dimensional general relativity recasted as a deSitter gauge theory. We proceeded by introducing local symmetry breaking terms in the action supported by worldlines and worldsheets. Restoration of full gauge invariance has led to the introduction of new dynamical fields interpreted as describing matter degrees of freedom. The diffeomorphism symmetry led us to the notion of position or embedding, while the local Lorentz symmetry gave rise to variables encoding momentum and spin.
This physical interpretation was established by relating our formalism to the description of particle and string theories in terms of Poincar\'e group coordinates, or pseudo-classical variables \`a la Balachandran and collaborators. We have then studied the variational problem of the coupled system. We have calculated the deformation of the momentum and total angular momentum conservation laws due to curved spacetime effects. This has led us to Mathisson-Papapetrou and spin precession equations both for particles and strings. We have then derived the equations describing the gravitational field produced by such matter sources. For the non-spinning string case, we have discussed some solutions related to cosmic strings. 
Finally, we have explained why this formulation of (spinless) strings was promising from the spinfoam quantum gravity perspective.

We believe that the framework presented here provides potentially interesting outcomes in various directions of research, the investigations of which are currently under study. 

\section*{Acknowledgements}

Thanks to Aristide Baratin, Alejandro Perez, John W. Barrett and Carlo Rovelli for discussions. 

This work was supported by the Royal Commission for the Exhibition of 1851.



\begin{thebibliography}{30}

\bibitem{Freidel1}
L. Freidel, D. Louapre, ``Ponzano-Regge model revisited I: Gauge fixing, observables,
and interacting spinning particles"
{\em Class. Quant. Grav.}, {\bf 21}:5685-5726 (2004), [arXiv:hep-th/0401076]

\bibitem{effqg}
  L.~Freidel,  E.~R.~Livine,
  ``Effective 3d quantum gravity and non-commutative quantum field theory'',
 {\it Phys.\ Rev.\ Lett.\ }, {\bf 96}:221301 (2006),
  [arXiv:hep-th/0512113];  
  
   L.~Freidel and E.~R.~Livine,
   ``Ponzano-Regge model revisited III: Feynman diagrams and effective  field
  theory'',
  {\it Class.\ Quant.\ Grav.\ }, {\bf 23}:2021 (2006),
  [arXiv:hep-th/0502106]

\bibitem{karim}
K. Noui, ``Three dimensional Loop Quantum Gravity: Towards a self-gravitating Quantum Field Theory'',
{\it Class. Quant. Grav.}, {\bf 24}:329-360 (2007),
[arXiv:gr-qc/0612145]

\bibitem{BFloops}
J. C. Baez, D. K. Wise, A. S. Crans, ``Exotic statistics for loops in 4-D BF theory'',
{\it Adv. Theor. Math. Phys.}, {\bf 11}:707-749 (2007),
[arXiv:gr-qc/0603085]

\bibitem{BP}
J. C. Baez, A. Perez, ``Quantization of strings and branes coupled to BF theory'', {\it Adv. Theor. Math. Phys.}, {\bf 11}:3 (2007),
[arXiv:gr-qc/0605087]

\bibitem{AW}
W. J. Fairbairn, A. Perez, ``Extended matter coupled to BF theory'', {\em Phys. Rev.}, {\bf D78}:024013 (2008),
[arXiv:gr-qc/0709.4235]

\bibitem{Ale}
M. Montesinos, A. Perez, ``Two-dimensional topological field theories coupled to four-dimensional BF theory'',
{\em Phys. Rev.}, {\bf D77}:104020 (2008)
[arXiv:gr-qc/0711.2875] 

\bibitem{Bal1}
A. P. Balachandran, G. Marmo, B. S. Skagerstam, A. Stern, ``Gauge symmetries and Fibre bundles, Applications to Particle Dynamics, {\em Lecture Notes in Physic}, 188 (1982)

\bibitem{Utiyama}
R. Utiyama, ``Invariant theoretical interpretation of interactions'', {\em Phys. Rev.}, {\bf 101}:1597 (1956)

\bibitem{Kibble}
T. W. B. Kibble, ``Lorentz invariance and the gravitational field'', 
{\em J. Math. Phys.}, {\bf 2}:212-221 (1961)

\bibitem{Sciama} 
D. W. Sciama, ``The physical structure of general relativity''
{\em Rev. Mod. Phys.}, {\bf 36}:463-469 (1964)

\bibitem{Leclerc}
M. Leclerc, ``The Higgs sector of gravitational gauge theories'', {\em Annals Phys.}, {\bf 321}:708-743 (2005), 
[arXiv:gr-qc/0502005]

\bibitem{Ellis}
G. J. Ellis, ``A non-abelian tensor product of Lie algebras'', {\it Glasgow Math. J.}, {\bf 33}:101 (1991)

\bibitem{desousa}
Ph. de Sousa Gerbert, ``On spin and (quantum) gravity in (2+1)-dimensions'',
{\it Nucl.Phys.} {\textbf B}346:440-472 (1990)

\bibitem{Pleb}
J. F. Plebanski, ``On the separation of Einstein substructures'', {\em J. Math. Phys.}, {\bf 12}:2511 (1977) 

\bibitem{Bal2}
A. P. Balachandran, F. Lizzi, G. Sparano, ``A new approach to strings and superstrings'', {\em Nucl. Phys.}, {\bf B277}:359 (1986) ;

A. P. Balachandran, F. Lizzi, R. D. Sorkin, G. Sparano, ``Topological Aspects Of String Theories'', 
{\em Nucl. Phys.}, {\bf B287}:508 (1987)

\bibitem{Finkelstein}
D. Finkelstein, ``Internal structure of spinning particles'', {\em Phys. Rev.}, {\bf 100}:924-931 (1955)

\bibitem{Stern}
A. Stern, ``Bosonization and Current Algebra of Spinning Strings'', 
{\em Nucl.Phys.}, {\bf B482}:305-324 (1996), [arXiv:hep-th/9610014] 

\bibitem{Stern1}
B. Harms, A. Stern, ``Bosonic description of spinning strings in (2+1)-dimensions'',
{\em Phys .Rev.}, {\bf D54}:3981-3984 (1996), 
[arXiv:hep-th/9601155]

\bibitem{mm}
S. W. MacDowell, F. Mansouri, ``Unified geometric theory of gravity and supergravity'', {\it Phys. Rev. Lett.}, {\bf 38}:739-742 (1977)

\bibitem{wise}
D. K. Wise, ``MacDowell-Mansouri gravity and Cartan geometry'' (2006), [arXiv:gr-qc/0611154]

\bibitem{artem}
L. Freidel, J. Kowalski-Glikman, A. Starodubtsev,
``Particles as Wilson lines of gravitational field'', {\it Phys. Rev.}, {\bf D74}:084002 (2006), [arXiv:gr-qc/0607014]

\bibitem{Mathison}
M. Mathisson, {\em Acta. Phys. Polon.}, {\bf 6}:225 (1937); 

A. Papapetrou, ``Spinning test particles in general relativity. 1'', {\em Proc. Roy. Soc. Lond.}, {\bf A 209}:248 (1951) 

\bibitem{deser}
S. Deser, R. Jackiw, ``Time travel?'', {\em Comments Nucl.Part.Phys.}, {\bf 20}:337-354 (1992), [arXiv:hep-th/9206094]

\bibitem{FP}
R. De Pietri, L. Freidel ``so(4) Plebanski action and relativistic spin foam model''
{\em Class. Quant. Grav.}, {\bf 16}:2187-2196 (1999), [arXiv:gr-qc/9804071]

\bibitem{SF}
C. Rovelli, {\em Quantum Gravity}, (Cambridge Univ. Pr.), (2004) ;

A. Perez, ``Spin foam models for quantum gravity'', {\em Class. Quant. Grav.},
{\bf 20}:R43 (2003), [arXiv:gr-qc/0301113];

D. Oriti, ``Spacetime geometry from algebra: Spin foam models for non-perturbative  quantum gravity'', {\em Rept. Prog. Phys.}, {\bf 64}:1489-1544 (2001), [arXiv:gr-qc/0106091];

J. C. Baez, ``An introduction to spin foam models of BF theory and
                  quantum gravity'',
{\em Lect. Notes Phys.}, {\bf 543}:25-94 (2000), [arXiv:gr-qc/9905087];

J. C. Baez, ``Spin foam models'', {\em Class. Quant. Grav.}, {\bf 15}:1827-1858 (1998),
[arXiv:gr-qc/9709052]

\bibitem{ARW}
W. J. Fairbairn, R. Pereira, ``Strings in spinfoam models'', to appear

\end{thebibliography}
\end{document}